\documentclass{article}
\usepackage{spconf,amsmath,graphicx}
\usepackage[hyphens]{url}
\usepackage{hyperref}
\ninept
\usepackage[table,xcdraw]{xcolor}

\title{Overview of the L3DAS23 Challenge on Audio-Visual Extended Reality}
\makeatletter
\def\@name{
\emph{Christian~Marinoni$^{\dagger}$}, \emph{Riccardo~F.~Gramaccioni$^{\dagger}$},
\emph{Changan~Chen$^{\ast}$},
\emph{Aurelio~Uncini$^{\dagger}$}, 
\emph{Danilo~Comminiello$^{\dagger}$}

\thanks{Corresponding author's email: \href{mailto:danilo.comminiello@uniroma1.it}{danilo.comminiello@uniroma1.it}. This work has been partly supported by ``Progetti di Ricerca'' of Sapienza University of Rome under grant numbers RG11916B88E1942F and RM120172AC5A564C, and by ``Centro Nazionale 1 - Spoke 6'' of Sapienza University of Rome under grant number CN1321845CE18353.}\vspace{1em}}
\makeatother
\address{$^{\dagger}$DIET Dept., Sapienza University of Rome, 
Italy \\ 
$^{\ast}$UT Austin, TX, USA
}

\begin{document}
%
\maketitle
\begin{abstract}
The primary goal of the L3DAS23 Signal Processing Grand Challenge at ICASSP 2023  is to promote and support collaborative research on machine learning for 3D audio signal processing, with a specific emphasis on 3D speech enhancement and 3D Sound Event Localization and Detection in Extended Reality applications. As part of our latest competition, we provide a brand-new dataset, which maintains the same general characteristics of the L3DAS21 and L3DAS22 datasets, but with first-order Ambisonics recordings from multiple reverberant simulated environments. Moreover, we start exploring an audio-visual scenario by providing images of these environments, as perceived by the different microphone positions and orientations. 
We also propose updated baseline models for both tasks that can now support audio-image couples as input and a supporting API to replicate our results. Finally, we present the results of the participants. Further details about the challenge are available at \texttt{\href{https://www.l3das.com/icassp2023}{www.l3das.com/icassp2023}}.
\end{abstract}
\section{Introduction}
\label{sec:intro}

3D audio applications, such as virtual and real conferencing, game development, music production, augmented reality and immersive technologies in virtual environments, are gaining increasing interest in the machine learning community in recent years.
In this context, the L3DAS22 Challenge \cite{9746872}, organized within the L3DAS (Learning 3D Audio Sources) project and presented as Grand Challenge at ICASSP 2022, introduced a dataset based on multiple-source and multiple-perspective (MSMP) Ambisonics recordings for two 3D audio tasks: 3D Speech Enhancement (SE) and 3D Sound Event Localization and Detection (SELD). More precisely, 3D SE aims to improve speech clarity and intelligibility by removing undesired information from noisy spatial vocal recordings. 3D SELD seeks spatiotemporal descriptions of 3D acoustic scenes, predicting the start and end times of sounds in the recording and their categories. Influenced by the recent growing interest in virtual and augmented reality (VR \& AR), the L3DAS23 Challenge proposes two substantial evolutions: (a) instead of being obtained from a real-existing environment, the 3D audio recordings that compose the dataset come from 68 different simulated environments; (b) it introduces a separate track for the multimodal scenario, combining audio recordings and RGB images. 
Therefore, participants can choose to use only the MSMP Ambisonics audio recordings or consider both audio and visual features, as described in Section \ref{sec:challenge_tasks}. We expect that the characterization of the environment can provide valuable support in the execution of the two tasks.
We supply baseline models for both tasks, two 3D audio datasets and a Python-based API that facilitates the data download and preprocessing, the baseline models training and the results submission.

\section{DATASET DESCRIPTION}
\label{sec:description}
\begin{figure}[htb]

%
\begin{minipage}[b]{.48\linewidth}
  \centering
  \centerline{\includegraphics[width=4.0cm]{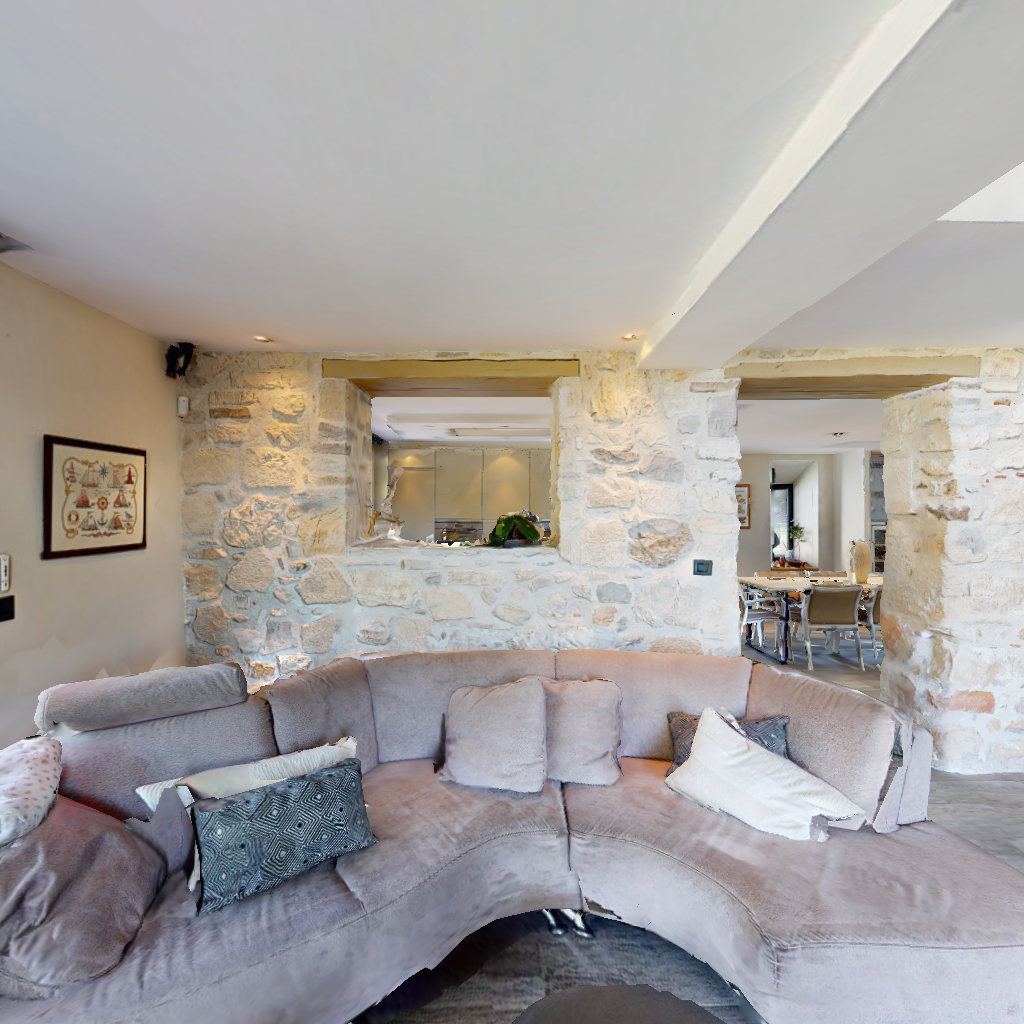}}
\end{minipage}
\hfill
\begin{minipage}[b]{0.48\linewidth}
  \centering
  \centerline{\includegraphics[width=4.0cm]{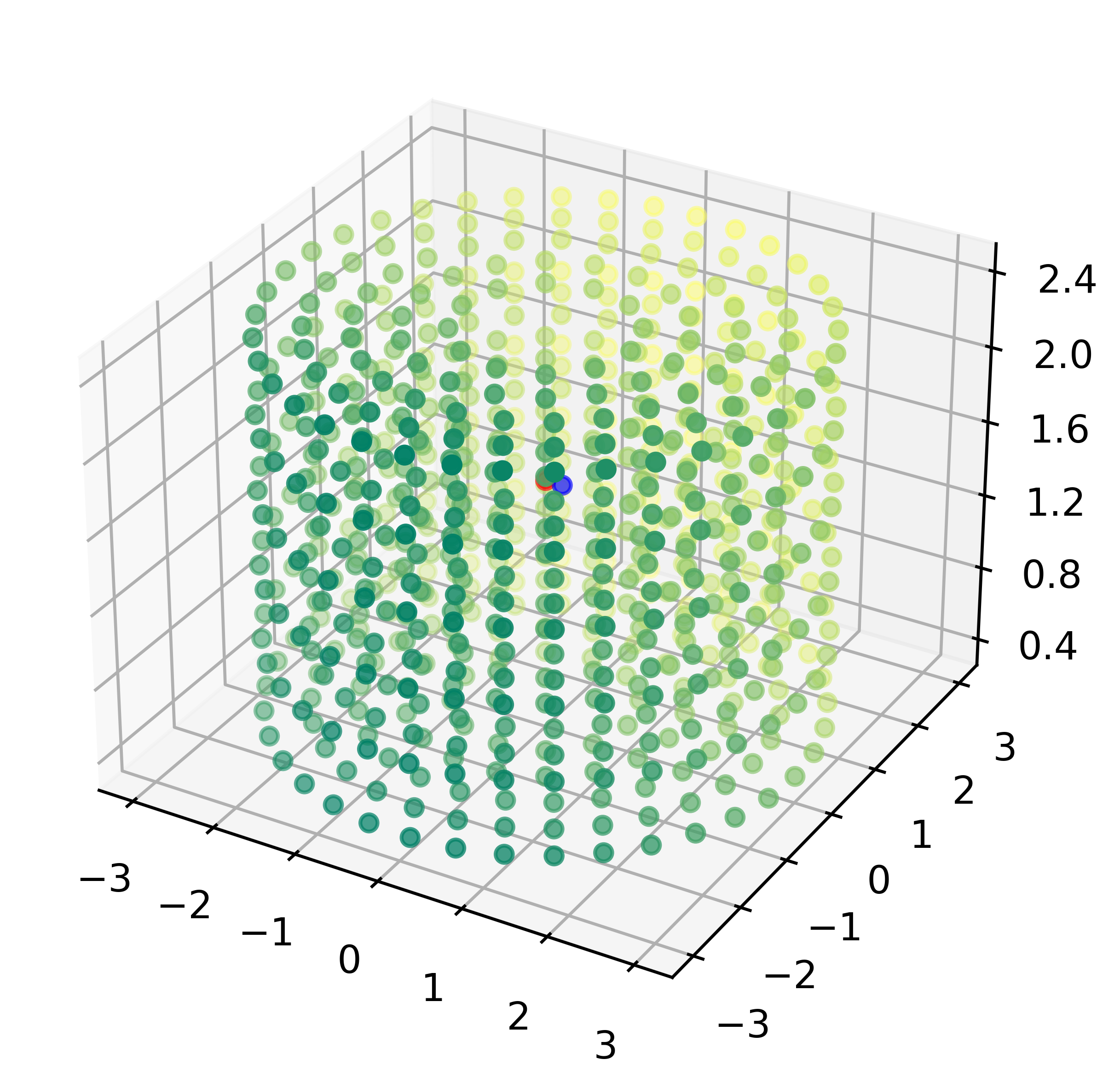}}
\end{minipage}
\caption{Example of simulated environment (left) and cylindrical grid used for sound positioning (right).}
\label{fig:example1}
\end{figure}

The L3DAS23 dataset\footnote{The dataset was built by employing an NVIDIA Quadro RTX 8000 thanks to the NVIDIA Applied Research Accelerator Program.} contains approximately 100 hours of MSMP B-format audio recordings.
We use Soundspaces 2.0 \cite{chen22soundspaces2} to generate both Ambisonic Room Impulse Responses (ARIRs) and images in a selection of simulated 3D houses from the Habitat - Matterport 3D Research Dataset \cite{ramakrishnan2021hm3d}.
Each simulated environment has a different size and shape and includes multiple objects and surfaces to which specific acoustic properties (i.e., absorption, scattering, transmission, damping) are applied.
We locate two Ambisonics microphones in random positions in each house (mic B is distant 20 cm towards the $x$ dimension from mic A) and generate B-Format ACN/N3D impulse responses of the room by placing the sound sources according to a cylindrical grid centred in mic A, as shown in Figure \ref{fig:example1}. Both microphones are at the same height of 1.6 m and have the same orientation.
Relying on the collected Ambisonics impulse responses, we augment existing clean monophonic datasets to obtain synthetic tridimensional sound sources by convolving the original sounds with the IRs extracted from the virtual environments. 
This convolution operation results in the virtual placement of a sound source in specific spatial positions of the house, as perceived from the two microphones. To create plausible and variegate 3D scenarios, we use Librispeech \cite{DBLP:conf/icassp/PanayotovCPK15} for the speech signals and 1440 noise sound files, divided into 14 transient noise classes and 4 continuous noise classes.
In addition to the Ambisonics recordings, the dataset provides, for each microphone position in the rooms, a 512x512 px image, representing the environment in front of the microphone. These images are obtained by virtually placing an RGB sensor at the same height and orientation as microphone A and with a 90-degree field of view.

\section{CHALLENGE TASKS}
\label{sec:challenge_tasks}
We propose two distinct tasks: 3D Speech Enhancement in Simulated Reverberant Environments and 3D Sound Event Localization and Detection in Simulated Reverberant Environments.
Each task includes two tracks: audio-only and audio-visual, respectively relying on sound recordings only or a combined use of audio and images. Finally, each track is split into two sub-tracks: one-mic and dual-mic recordings, respectively proposing the sounds acquired by one or both Ambisonics microphones.

\subsection{Task 1: 3D Speech Enhancement}
\label{subsec:SE}
Task 1 includes over 40,000 virtual 3D audio recordings, each lasting up to 12 seconds, resulting in approximately 90 hours of total duration.
The main objective of this task is to separate and enhance speech signals immersed in noisy virtual environments. Hence, the models should extract the monophonic voice signal from the 3D mixture that contains various background noises. The predictors are 8-channel 16 kHz 16-bit wav files (2 sets of first-order Ambisonics recordings) and, optionally, RGB image files. The target data provided for this section contains the clean monophonic recordings of the speech signals (16 kHz 16-bit mono wav files) with corresponding words in a CSV file.
The evaluation metric for this task combines Short-Time Objective Intelligibility (STOI) and Word Error Rate (WER) to assess the effect of enhancement on speech recognition.
The final metric for this task is a combination of these two measures given by \((STOI+(1-WER))/2\).
We use a Wav2Vec architecture pre-trained on Librispeech 960h to compute the WER.
For Task 1, the baseline we provide is a beamforming U-Net architecture as the one in L3DAS22, which we also present in a new variant (namely, audiovisual MIMO UNet Beamforming) that concatenates the visual features with the high-level features generated as output of the encoder part of the U-Net. Our best model scores a T1 metric of 0.553, with a WER of 0.567 and a STOI of 0.673.

\subsection{Task 2: 3D Sound Event Localization and Detection}
\label{subsec:SE}
Task 2 aims to identify and locate some known sound events in a synthetic 3D acoustic environment. We generate 900 30-second-long data points (a total of 7.5 hours), having up to 3 acoustic events simultaneously active while guaranteeing that concurrent sounds of the same class always have a minimum linear distance of 1 meter from one another.
The data used for prediction has the same format as the SE section, except for the sampling frequency, which is 32 kHz.
As target data, we provide a CSV file containing the onset and offset time stamps, the class and the spatial coordinates of each sound event present in a data point. 
Here we employ a unified metric for detection and localization, known as Location-sensitive Detection Error. This metric measures the Cartesian distance between the predicted and true events and only considers a true positive when the label is correct and the location is within a threshold (1.75 m) of the reference location.
For Task 2, we extend the L3DAS22 baseline to the audio-visual scenario by extracting visual features and concatenating them to the audio features right before the two network heads. Our best model achieves a Task 2 metric of 0.158.

\section{Challenge results}
Overall, 4 teams participated in Task 1 and 2 in Task 2. Tables 1 and 2 show the results obtained by the participants on the test set. Special mention to the JLESS team, which participated in both tasks and proposed an audio-visual version of its model for Task 2. The estimated carbon footprint of the whole competition is approximately 110 kg. CO2 eq.

\begin{table}[]
\centering
\begin{tabular}{ccccc}
\rowcolor[HTML]{FFDA9C} 

Rank & Team Name                      & WER   & STOI  & T1 Metric \\ \hline
1    & \multicolumn{1}{l}{SEU Speech} & 0.101 & 0.902 & 0.901     \\
2    & JLESS                          & 0.174 & 0.836 & 0.831     \\
3    & CCA Speech                     & 0.240 & 0.831 & 0.796     \\
-    & Baseline                       & 0.567 & 0.673 & 0.553     \\
4    & SpeechLab410                   & 0.643 & 0.608 & 0.483    
\end{tabular}
\caption{Results of Task 1 participants.}
\label{tab:task1}
\end{table}

\begin{table}[]
\centering
\begin{tabular}{ccccc}
\rowcolor[HTML]{FFDA9C} 
Rank* & Team Name     & Precision   & Recall  & T2 Metric \\ \hline
1    & JLESS         & 0.288 & 0.204 & 0.239     \\
2    & NERCSLIP-USTC & 0.275 & 0.216 & 0.242     \\
-    & Baseline      & 0.182 & 0.140 & 0.158    
\end{tabular}
\caption{Results of Task 2 participants. *The final ranking takes into account the quality of the results produced, in addition to the T2 metric values on the test set.}
\label{tab:task2}
\end{table}

\section{Conclusion}
This paper includes an overview of the ICASSP 2023 Signal Processing Grand Challenge named L3DAS23 and a general outline of the two related datasets.
The L3DAS Team's future work will incorporate new 3D acoustic scenarios, diverse microphone configurations, and novel tasks toward more complete audio-visual setups. We also count on tuning the challenge complexity and overcoming some limitations of the current edition.



\bibliographystyle{IEEEbib}
\bibliography{strings,refs}

\end{document}